\documentclass[lettersize,journal]{IEEEtran}
\usepackage{amsmath,amsfonts,algorithmic,algorithm,array,textcomp,stfloats,url,verbatim,cite,tikz,amsmath,xspace,mathrsfs,tcolorbox,xcolor,graphicx,float,filecontents,comment,listings,booktabs,caption,bm,subcaption,multirow, enumitem}

\PassOptionsToPackage{hyphens}{url}\usepackage{hyperref}
\newcommand{\mypar}[1]{\noindent\textbf{#1}}

\newcommand{\obelisk}{OBELISK\xspace}

\newcommand{\standard}{STND\xspace}
\newcommand{\gbdtall}{GBDT-ALL\xspace}
\newcommand{\gbdtfilter}{GBDT-YARA\xspace}
\newcommand{\nebulaall}{NEBULA-ALL\xspace}
\newcommand{\nebulafilter}{NEBULA-GBDT-YARA\xspace}
\newcommand{\obvone}{OBV1\xspace}
\newcommand{\obvtwo}{OBV2\xspace}
\newcommand{\obvthree}{OBV3\xspace}

\usepackage[textwidth=2cm,textsize=tiny]{todonotes}

\newlength{\thmlabelwidth}
\begin{document}
\title{\Large \bf Windows Malware Detector as a Compound AI System:\\
Trade-Offs in Accuracy, Efficiency, and Adversarial Robustness}
\author{Andrea Ponte, Luca Demetrio,~\IEEEmembership{Member,~IEEE},\\
Luca Oneto,~\IEEEmembership{Senior~Member,~IEEE}, Battista Biggio,~\IEEEmembership{Fellow,~IEEE}, Fabio Roli,~\IEEEmembership{Fellow,~IEEE}
\thanks{Andrea Ponte, Luca Demetrio and Luca Oneto are with the Department of
Informatics, Bioengineering, Robotics and Systems Engineering, University of
Genova, 16126 Genova, Italy (e-mail: andrea.ponte@edu.unige.it).}
\thanks{Battista Biggio is with the Department
of Electrical and Electronic Engineering, University of Cagliari, 09123 Cagliari,
Italy.}
\thanks{Fabio Roli is with the Department of
Informatics, Bioengineering, Robotics and Systems Engineering, University of
Genova, 16126 Genova, Italy, and also with the Department
of Electrical and Electronic Engineering, University of Cagliari, 09123 Cagliari,
Italy.}}
\maketitle
\begin{abstract}
Industrial Windows malware detectors are commonly described as Compound AI Systems composed of multiple heterogeneous components, including rule-based mechanisms as well as machine-learning-based static and dynamic analyses.
However, due to industrial secrecy and limited public disclosure, the internal architectures of these systems can only be inferred, rendering systematic evaluations of detection accuracy, computational costs, and adversarial robustness largely infeasible.
In contrast, academic research provides reproducible and transparent evaluation methodologies, but typically investigates individual detection components in isolation.
To bridge the gap between academic research and industrial practice, and inspired by state-of-the-art industrial architectures for Windows malware detection, we propose a novel methodology that (i) explicitly balances the trade-off among detection performance, computational requirements, and robustness, and introduces (ii) system-level threat models that capture how attackers exploit different degrees of knowledge to evade the entire Compound AI System rather than isolated detectors.
Experiments conducted on real-world data demonstrate that the Compound AI System training time can be reduced and responsiveness improved while incurring only a marginal loss in detection performance.
Leveraging our threat modeling, we show that increasingly knowledgeable attackers craft more effective adversarial examples, revealing the system's strengths and weaknesses,
degrading its responsiveness, and exposing a direct trade-off between efficiency and robustness.
Finally, we translate these trade-offs into take-home messages and deployment guidelines, helping practitioners to select the system that best matches their operational constraints.
\end{abstract}
\begin{IEEEkeywords}
Windows Malware Detection, Compound AI System, Accuracy, Efficiency, Threat Models, Robustness.
\end{IEEEkeywords}
\section{Introduction}
\label{introduction}
When browsing antivirus solutions to install or purchase, most products emphasize the adoption of multi-layered detection pipelines.
These typically combine traditional detection techniques with advanced Machine Learning (ML) models~\cite{crowdstrikefalcon_whitepaper, kaspersky_whitepaper, avast_blog, windowsdefender_blog, bitdefender_whitepaper}, often collectively referred to as \emph{Compound AI Systems}.
Due to industrial secrecy and limited public disclosure, the internal structure of these Compound AI Systems can only be partially inferred.
They are commonly assumed to operate as sequential pipelines that trade detection effectiveness for computational efficiency~\cite{windowsdefender_blog, kaspersky_whitepaper, crowdstrikefalcon_whitepaper}.
Such pipelines usually begin with a signature-based detector, which is deployed for its reliability in identifying well-known malware samples~\cite{apple_xprotext, kaspersky_rules}.
This stage is typically followed by ML models trained on static analysis, i.e., inferring maliciousness from code and metadata without executing the program~\cite{raff2017malware, raff2021classifying, coull2019activation, krvcal2018deep}.
Finally, the last line of defense is provided by models trained on dynamic analysis, which infer maliciousness from runtime behavior observed when samples are executed in isolated virtual environments~\cite{trizna2024nebula, ling2022malgraph, jindal2019neurlux}.
These dynamic analysis components may be deployed locally or executed in the cloud.
However, the secrecy surrounding these pipelines renders systematic and transparent evaluations infeasible.
While competitive benchmarking platforms exist\footnote{\url{https://www.av-test.org/en/antivirus/business-windows-client/}} and report metrics such as detection rates and false positive counts on closed sets of malicious and benign samples, they do not enable comparisons of Compound AI Systems along other critical dimensions, particularly overlooking computational requirements and robustness. 
As a result, it remains unclear whether - and to what extent - these systems can be evaded by carefully crafted and minimally modified programs, commonly referred to as \emph{adversarial EXEmples}~\cite{demetrio2021adversarial}.
In contrast, academic research offers systematic and transparent evaluations, but solely focused on isolated components rather than fully integrated Compound AI Systems~\cite{croce2robustbench, demetrio2021functionality}.
At design time, most research concentrates on training individual ML models within narrowly defined settings, either exclusively for static~\cite{raff2017malware, raff2021classifying, coull2019activation, krvcal2018deep}, dynamic~\cite{trizna2024nebula, ling2022malgraph, jindal2019neurlux}, or hybrid~\cite{trizna2022quo} analysis.
Alternatively, some works stack pre-trained models to emulate sequential detection pipelines~\cite{ponte2025slifer}.
At deployment time, due to the lack of system-wide security evaluations and formalized threat models for Compound AI Systems, research continues to focus almost exclusively on isolated models~\cite{kolosnjaji2018adversarial, demetrio2021adversarial, demetrio2021functionality, lucas2021malware, song2022mab, digregorio2024tarallo}.
In fact, at the time of writing, the only available approach to assess the robustness of Compound AI Systems for Windows malware detection consists of generating attacks against a single ML model and subsequently testing them against the real target system~\cite{demontis2019adversarial, demetrio2021functionality, ponte2025slifer, ponte2025demystifying}.
Consequently, academia has largely overlooked rigorous analyses of integrated systems, contributing to a growing disconnect between academic research and industrial practice.
%This gap manifests at every stage of the life cycle of modern anti-malware solutions.
%
To the best of the authors’ knowledge, there is currently no work systematically investigating the joint optimization of performance, computational requirements, and robustness while accounting for both development and deployment costs. 
%of Compound AI Systems for Windows malware detection.
Hence, in this paper, we aim to narrow the gap between industry and academia by proposing a novel methodology for optimizing and evaluating Compound AI Systems for Windows malware detection.
Our approach explicitly highlights the inherent trade-offs among detection performance, computational requirements (both at design and deployment stages), and robustness that arise from integrating multiple detection technologies within a single pipeline (\autoref{sec:sys_form}).
%With respect to pipeline optimization, we focus on the trade-off between performance and computational cost at both the design and deployment stages.
%
During the design stage, the primary computational costs stem from training the ML models for static and dynamic analysis.
Reducing the amount of training data by exploiting the fact that a subset of samples is filtered out at earlier pipeline stages directly lowers these costs.
However, using less training data may negatively impact performance, as reduced data availability often leads to lower model accuracy and may also hinder robustness.
In this work, we identify when it is possible to reduce the training data while limiting the impact on performance, and when such reductions become detrimental.
At the deployment stage, the pipeline components are naturally ordered by increasing computational cost: deterministic signature-based detectors, which incur negligible overhead (hundredths of a second); probabilistic static ML models, which require slightly higher computation times (fractions of a second); and probabilistic dynamic ML models, which are significantly more expensive, often requiring several seconds per sample.
We show that optimizing the probabilistic thresholds used to decide whether to trigger more computationally expensive analyses allows one to jointly optimize performance and computational cost.
With respect to robustness evaluation, we first formalize a range of threat models under varying assumptions about the attacker’s knowledge of the system (\autoref{sec:robustness}).
We begin with a zero-knowledge adversary, who has no information about the system, and progressively consider stronger attackers who are aware of its structure but possess detailed knowledge of only a subset of its components.
We then restrict our analysis to threat models that are feasible in practice, given the current state of the art in attack libraries targeting individual components or entire Compound AI Systems.
Specifically, we focus on manipulations that affect only parts of the pipeline (e.g., the static ML component and/or signature-based detectors), as no effective manipulations are currently available for the remaining components.
We conduct a series of experiments on real-world data using \obelisk, our implementation of a Compound AI System for Windows malware detection (\autoref{sec:internals}) inspired by recent work~\cite{ponte2025slifer, ponte2025demystifying}, which is optimized and evaluated according to our proposed methodology (\autoref{sec:experiments}).
Our results show that the proposed filtering mechanism significantly reduces resource consumption while incurring only a marginal decrease in performance (\autoref{sec:performance}).
Moreover, our system-level threat models reveal that better-informed attackers can mount more effective evasion attacks, degrading system responsiveness and exposing the system's strengths and weaknesses that remain hidden when detectors are evaluated in isolation (\autoref{sec:adv_evaluation}). 
Also, we show that the same filtering mechanism used to reduce computational cost simultaneously widens the system's attack surface and, by training on smaller datasets, further weakens robustness, exposing a direct trade-off also between computational savings and robustness (\autoref{sec:levels}).
Finally, we distill our findings into a set of take-home messages and practical guidelines (\autoref{sec:recommendations}), to help practitioners select the configuration that best matches their operational constraints.

\section{Background and Related Work}
\label{sec:background}
\begin{figure}[t]
\centering
\includegraphics[width=0.85\linewidth]{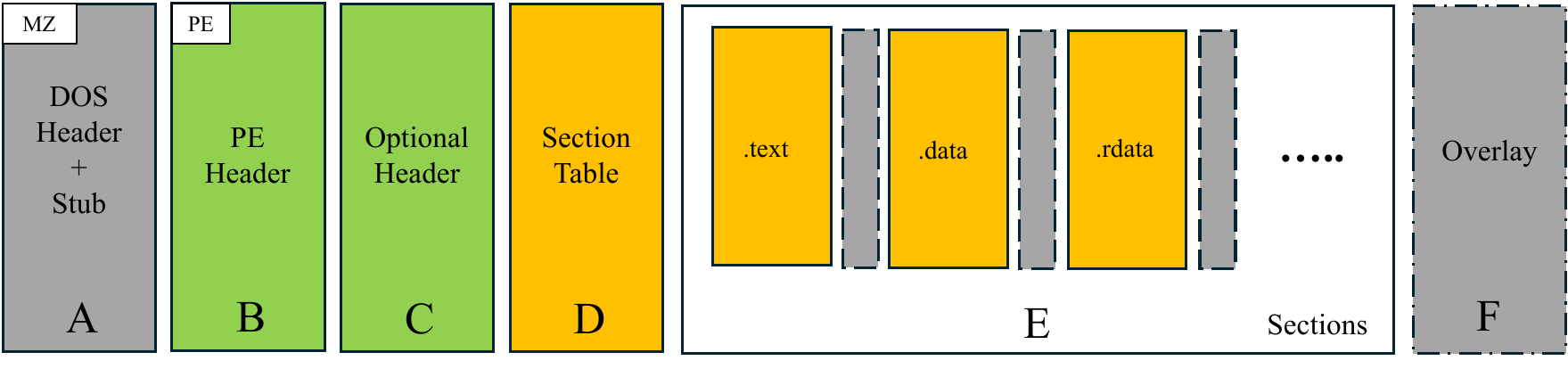}
\caption{Depiction of the Windows PE file format.}
\label{fig:pe}
\end{figure}
A Windows program is represented on disk as a file conforming to the Portable Executable (PE) format\footnote{\label{format}\url{https://learn.microsoft.com/en-us/windows/win32/debug/pe-format}} (\autoref{fig:pe}).
The main components of the PE format are summarized as follows: (A) the DOS header and stub, a valid DOS program retained for backward compatibility; (B) the PE header, which contains metadata describing the program; (C) the Optional Header, which stores information required for loading and executing the program; (D) the Section Table; and (E) the Sections, which contain the actual program content (e.g., the ``.text'' section contains compiled code).
We refer to those Windows malware samples that are crafted to evade ML-based detectors~\cite{demetrio2021adversarial} as \emph{adversarial EXEmples}.
These samples are generated through the application of \emph{practical manipulations}, i.e., modifications to the structure of input programs that preserve their original functionality and do not cause errors at load time.
Such manipulations exploit blind spots and redundancies in the PE file format, allowing attackers to insert, remove, or replace content without compromising the correctness of the executable.
For example, attackers can manipulate and extend headers (gray area in \textbf{A}~\cite{demetrio2021adversarial}), inject new sections~\cite{demetrio2021functionality}, fill unused space between sections (gray areas in \textbf{E}~\cite{kreuk2018deceiving}), or append arbitrary bytes at the end of the last section (gray area in \textbf{F}~\cite{kolosnjaji2018adversarial}).
The content of these manipulations is typically determined by an optimization algorithm, which may operate in a \emph{black-box} setting - assuming no knowledge of the detector’s internal structure - via \emph{query-based} or \emph{transfer} attacks, or in a \emph{white-box} setting - assuming full knowledge of the detector - via gradient-based optimization methods if the detector is differentiable~\cite{kolosnjaji2018adversarial, lucas2021malware}.
We are currently witnessing a growing gap between academic research and industrial practice.
On the one hand, industry increasingly relies on complex detection solutions composed of multiple interacting components~\cite{crowdstrikefalcon_whitepaper, kaspersky_whitepaper, avast_blog, windowsdefender_blog, bitdefender_whitepaper}. %,commonly referred to as \emph{Compound AI Systems}.
On the other hand, academic research has largely focused on the design and evaluation of individual models, typically based on static, dynamic, or hybrid analysis~\cite{raff2017malware, coull2019activation, jindal2019neurlux, trizna2024nebula, trizna2022quo, chaganti2023multi}.
To the best of our knowledge, only a limited number of works have explicitly investigated the properties of Compound AI Systems for Windows malware detection~\cite{ponte2025slifer, ponte2025demystifying}.
While relevant, these contributions represent only an initial step toward a systematic study of the trade-offs among accuracy, efficiency, and adversarial robustness in AI-based malware detection.
In contrast to this work, prior studies do not (i) explicitly address the trade-off between performance, computational requirements, and robustness or (ii) define and evaluate system-level threat models that capture how attackers exploit different degrees of knowledge to evade a Compound AI System.
\section{Compound AI Systems for Windows Malware Detection}
\label{sec:methodology}
\begin{figure*}[t]
\centering
\includegraphics[width=0.82\linewidth]{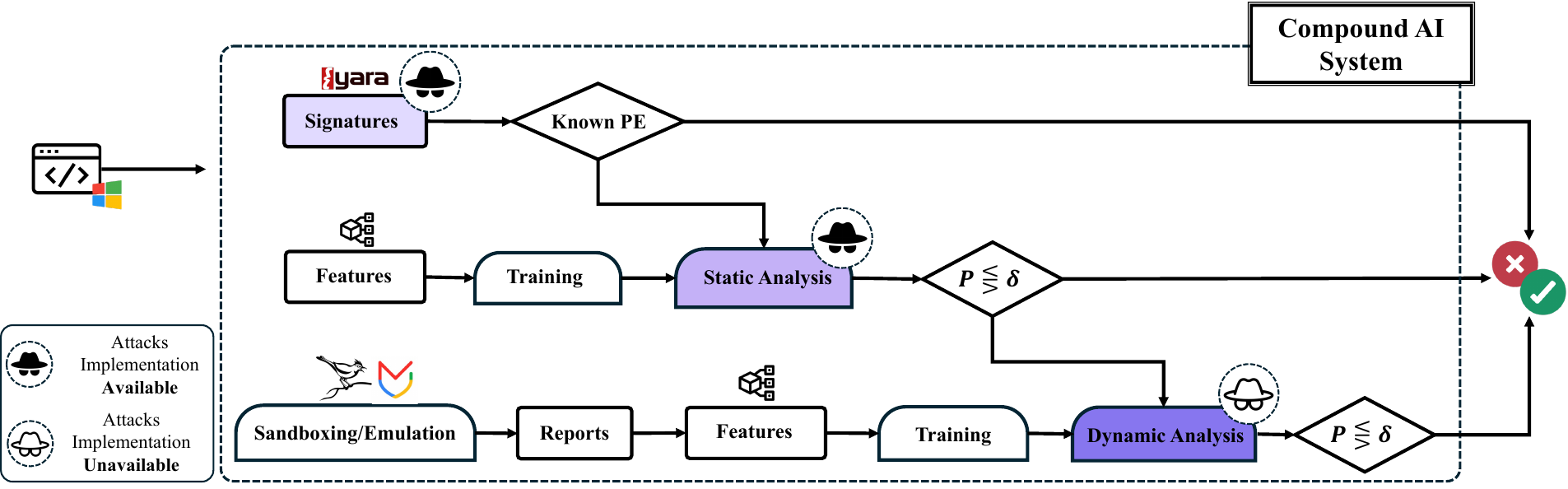}
\caption{Considered Compound AI System for Windows malware detection based on a sequential architecture.
Each analysis stage is characterized by its intermediate pre-processing and deployment steps, and the criteria used to forward samples through the pipeline.
All components are annotated with the presence or absence of known attacks reported in the literature, as well as with the availability of open-source, stable tools that implement such attacks for assessing the robustness of the ML models.}
\label{fig:concept}
\end{figure*}

To bridge the gap between academic research and real-world deployments, it is necessary to move beyond the prevailing assumption of isolated ML components acting as standalone defense mechanisms.
Instead, we advocate a \emph{Compound AI System} paradigm, in which multiple ML components cooperate and are complemented by conventional software responsible for automation and pre-processing, with the explicit goal of trading off detection performance and computational requirements (\autoref{sec:sys_form}).
This shift must be accompanied by comprehensive threat modeling that considers varying assumptions about the attacker’s knowledge of the detection pipeline, explicitly accounts for interdependencies among components, and focuses on attacks that are practically feasible given the current state of the art in available tools (\autoref{sec:robustness}).
\subsection{Compound AI System Paradigm}
\label{sec:sys_form}
Based on the state of the art~\cite{windowsdefender_blog, kaspersky_whitepaper, crowdstrikefalcon_whitepaper}, real-world implementations of Compound AI Systems for Windows malware detection typically adopt a three-layer architecture:
(i) signature-based pattern matching;
(ii) static analysis of sample metadata; and
(iii) behavioral profiling via dynamic analysis, as illustrated in \autoref{fig:concept}.
This ordering reflects a deliberate trade-off between the use of domain knowledge - encoded as fast-to-evaluate, rigid logical constraints (i.e., signatures) - and purely data-driven methods that rely on softer decision boundaries learned from computationally expensive features.
The layered design enables early filtering of samples that can be classified with high confidence at minimal computational cost, reserving more complex and resource-intensive analyses for cases in which prior knowledge is insufficient.
In the absence of computational constraints, such pipelines would naturally collapse into an end-to-end approach.
However, a layered architecture introduces dependencies across levels, whereby each component acts as a filter for subsequent ones by progressively reducing and refining the data passed downstream.
This mechanism can be advantageous when deploying relatively ``shallow'' models, which often benefit from smaller but highly representative datasets, especially when features encode domain expertise.
By contrast, aggressive filtering may be detrimental for deep neural networks, which are designed to learn representations directly from data and therefore require large training sets to reduce reliance on handcrafted features.

\mypar{Pattern-matching with Signatures.}
This level identifies well-known malicious and benign samples via pattern matching against previously-extracted knowledge, often deployed as a primary (if not the sole) line of defense~\cite{apple_xprotext, windowsdefender_blog, clamav}, as shown in \autoref{fig:concept}.
Such mechanisms are explicitly documented in commercial products, including Apple’s native macOS antivirus~\cite{apple_xprotext}, Bitdefender~\cite{bitdefender}, Microsoft Defender~\cite{windowsdefender_blog}, and CrowdStrike Agent~\cite{crowdstrike_rules}.
They are typically implemented as \emph{blocklists}, i.e., collections of byte patterns derived from reverse engineering malware samples; a match to any signature results in an immediate malware verdict, unlike other domains where rules often contribute to a soft anomaly score~\cite{scano2024modsec}.
Blocklists are commonly complemented by \emph{allowlists}, i.e., sets of hash values identifying known benign samples, which are essential to mitigate false alarms that may cause financial and reputational damage\footnote{\url{https://corelight.com/resources/glossary/false-positives-cybersecurity}}.
Signature-based detectors are simple to deploy, as updates only require modifying byte patterns, but they do not natively support tuning the trade-off between detection rate and false alarms.
Such tuning can be achieved either by (i) training a classifier that incorporates signature matches as features~\cite{scano2024modsec, gupta2024living}, at the cost of losing negligible inference overhead and effectively shifting signatures to the ML-based level, or (ii) removing signatures that produce excessive false positives.
During development, signatures can constrain the data distribution passed to downstream components, reducing their training requirements~\cite{ponte2025demystifying}.
After deployment, they immediately block known samples and forward only those lacking prior knowledge, typically including previously unseen samples produced via \emph{obfuscation techniques} that preserve functionality while altering the byte-level representation\footnote{\url{https://www.darkreading.com/threat-intelligence/only-half-of-malware-caught-by-signature-av}}.

\mypar{Static Analysis on Metadata.}
This level leverages program representations to assess whether a sample is malicious using a dedicated ML model, which requires an explicit feature extraction phase (see \autoref{fig:concept}).
Features may be manually engineered using domain knowledge and aggregation techniques~\cite{anderson2018ember}, or learned end-to-end~\cite{raff2017malware, coull2019activation} by exploiting structural dependencies in the data.
This enables highly effective detectors that extract knowledge directly from data, reducing reliance on time-consuming - though precise - manual reverse engineering.
Once features are defined, developers must (i) train the model on labeled benign and malicious samples and (ii) deploy it by specifying how samples are forwarded to subsequent pipeline stages.
Training is computationally expensive, whereas inference is relatively cheap; both can be automated and accelerated via increased computational resources, including GPUs at inference time.
Unlike signature-based detectors, which fundamentally rely on human expertise and manual updates, ML-based approaches scale primarily with available hardware.
Training for static analysis can be performed either (i) in isolation, ignoring upstream filtering and leveraging a larger but potentially less representative dataset, or (ii) jointly with the signature-based level, reducing data volume and computational cost while better matching the distribution observed after upstream filtering~\cite{ponte2025demystifying}.
At deployment time, the designer must decide how to propagate samples to the dynamic analysis stage, which is typically resource-intensive.
Unlike signatures, which produce binary decisions, this propagation is governed by the probabilistic output of the model.
Decision thresholds determine whether a prediction is confident enough to terminate the pipeline with a benign or malicious label, or whether the sample should be forwarded for further analysis, trading uncertainty for increased computational cost.
The same thresholding mechanism can be used during training by defining two cutoffs: samples below a lower threshold are labeled benign, those above an upper threshold malicious, and those in between are forwarded as training data to the next level.
While this strategy introduces a potential attack surface - since adversaries may attempt to keep samples below the lower threshold - it represents an explicit trade-off against system-wide computational constraints that must be considered by designers.

\mypar{Dynamic Analysis of Behavior.}
This level analyzes the runtime behavior of input samples, which is necessary because static representations cannot fully capture execution dynamics (see \autoref{fig:concept}).
In practice, malware may download additional payloads, inject code into other processes, unpack components in memory, or perform actions observable only at runtime.
Behavioral analysis is therefore performed by executing samples in a sandbox, i.e., an instrumented environment that allows programs to manifest their behavior.
While sandboxing can, in principle, provide the most comprehensive behavioral view, observing all relevant actions within a limited time budget is often infeasible.
This is due to (i) the need for multiple executions to explore different control-flow paths and (ii) deliberate evasion techniques whereby malware detects sandbox environments and suppresses or delays malicious behavior~\cite{afianian2019malware}.
Additionally, sandboxes enforce strict execution time limits to contain computational costs~\cite{kuchler2021does}.
Collected runtime traces are serialized into structured reports (e.g., JSON) describing API calls, network activity, and file-system operations.
Unlike static features, behavioral reports are difficult to encode with fixed, handcrafted feature sets and are therefore commonly processed using Natural Language Processing (NLP) techniques~\cite{trizna2024nebula, jindal2019neurlux}.
The resulting pipeline typically includes: (i) cleaning, to remove noisy or sample-specific fields (e.g., hashes); (ii) feature extraction, either learned directly from data or obtained via pre-trained models; and (iii) classifier training.
Both feature learning and classification are computationally expensive, as they rely on complex neural architectures whose effectiveness scales with dataset size.
These costs can be mitigated by using upstream levels as filters, reducing the number of samples processed at this stage and saving both training time and inference latency, with only a limited impact on detection performance (\autoref{sec:experiments}).
However, such filtering introduces a trade-off with representation expressiveness.
Behavioral modeling is inherently challenging because (i) individual executions generate large volumes of low-level events, often dominated by noise; (ii) benign and malicious programs may exhibit similar behaviors for different purposes; (iii) malware frequently employs obfuscation; and (iv) aggressive pre-processing may map distinct behaviors to similar feature representations.
Addressing these issues typically requires very large training datasets, increasing the overall computational burden.
After training, the model output is calibrated via decision thresholds that balance detection performance and false alarms.
This step is particularly critical, as dynamic analysis constitutes the final stage of the Compound AI System, where decisions are definitive.
Although dynamic analysis often underperforms static analysis when considered in isolation~\cite{dambra2023decoding}, it acts as a last line of defense against samples that evade earlier detectors.
Finally, independently of automated decisions, a subset of samples is usually analyzed manually by human experts to identify novel threats, variants of known malware, or benign programs not adequately captured by existing models (outside the scope of this work).

\mypar{Error Management.}
During the development and deployment of Compound AI Systems, practitioners must account for the possibility that any upstream level may fail.
For instance, a sandbox may crash on specific API calls, or a static pre-processing module may be unable to extract features due to malformed inputs.
Since the system must always return a decision, developers must define how such failures are labeled, explicitly accepting the corresponding trade-offs between increased false alarms and reduced detection rates~\cite{ponte2025slifer}.

\subsection{Evaluating Robustness of Compound AI Systems.}
\label{sec:robustness}
\begin{figure}[t]
\centering
\includegraphics[width=0.8\linewidth]{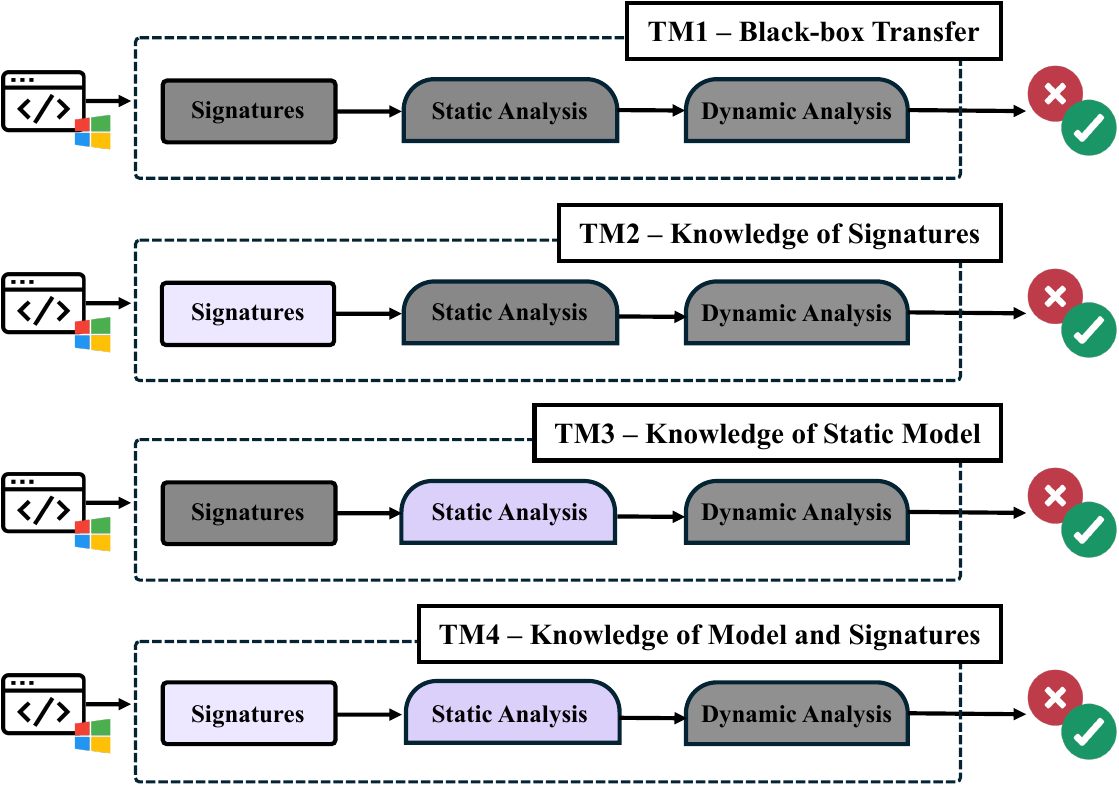}
\caption{The four threat models we consider, from the black-box one (TM1), i.e., the attacker is unaware of the components of the system, to the acquisition of partial knowledge  (TM4), i.e., the attacker knows how the first levels are implemented.}
\label{fig:tms}
\end{figure}
Isolated evaluations fail to capture the robustness of a Compound AI System as a whole.
They may \emph{underestimate} security, because upstream components can pre-process inputs and attenuate adversarial perturbations before they reach downstream detectors~\cite{ponte2025demystifying, ponte2025slifer}.
Conversely, they may \emph{overestimate} security, as adversaries can focus on the subset of samples that consistently evade the entire pipeline rather than individual components.
Robustness evaluations must hence explicitly account for system-level complexity by considering multiple threat models.
A natural starting point is a \emph{zero-knowledge} (black-box) evaluation~\cite{biggio2018wild}, in which attackers interact with the system only through its outputs, without access to internal details.
This setting includes:
(i) \emph{transfer-based attacks}, which assess robustness using adversarial samples crafted against a surrogate system with similar characteristics~\cite{demontis2019adversarial};
(ii) \emph{query-based attacks}, which iteratively optimize evasive samples by placing the target system in a feedback loop~\cite{demetrio2021functionality}, assuming queries are not severely constrained; and
(iii) obfuscation and packing techniques that alter representations while preserving functionality.
The latter are widely used in practice but are treated as a separate problem and are not addressed in this work.
Due to the high computational cost of querying the full pipeline, where inference on a single sample may take several seconds (as we will show in \autoref{sec:performance}), we focus on transfer-based evaluations.
The effectiveness of black-box attacks can be further increased by acquiring partial knowledge of the target via reverse engineering or information leakage, leading to a \emph{limited-knowledge} (gray-box) threat model that has proven effective even against commercial systems\footnote{\url{https://www.kb.cert.org/vuls/id/489481/}}.
At the other extreme, a \emph{perfect-knowledge} (white-box) attacker has full access to the system, a setting that, while potentially unrealistic, provides insight into worst-case vulnerabilities.
Accordingly, we define a spectrum of intermediate threat models - characterized by attacker goals, knowledge, and capabilities - in which the attacker's knowledge increases, while the goal (evasion) and capabilities (functionality-preserving manipulations at test time) remain consistent across models.

\mypar{Threat Model 1: Black-box Transfer (TM1).}
As a practical and widely adopted zero-knowledge setting, we consider an attacker with no access to the internals of the target Compound AI System (first row in \autoref{fig:tms}, unknown components in dark gray).
The attacker builds or collects a surrogate model with behavior similar to the target and crafts adversarial samples against it, which are then transferred to the real system, following established evaluation practices~\cite{demetrio2021functionality}.

\mypar{Threat Model 2: Knowledge of Signatures (TM2).}
Here, the attacker is aware of the presence of signature-based detection and can extract the deployed signatures, but ignores the existence of additional detection layers (second row in \autoref{fig:tms}).
This assumption is realistic, as signatures are often deployed locally~\cite{apple_xprotext, kaspersky_rules, bitdefender_whitepaper, windowsdefender_blog}.
The attacker can therefore optimize attacks to avoid artifacts detectable by signatures~\cite{ponte2025slifer, ponte2025demystifying}.

\mypar{Threat Model 3: Knowledge of Static Model (TM3).}
We assume the attacker can extract the static ML model used in the pipeline, but is unaware of any preceding signature-based filtering (third row in \autoref{fig:tms}).
This scenario is plausible given that vendors deploy ML models on endpoints~\cite{windowsdefender_blog}.
The attacker can perform worst-case (white-box) evaluations~\cite{demetrio2021adversarial} and test the resulting adversarial EXEmples against the full system.
We assume the attacker steals only the model parameters, not the decision thresholds, and therefore selects a reasonable operating point (1\% FPR) on an external dataset.

\mypar{Threat Model 4: Knowledge of Signatures and Static Model (TM4).}
In this setting, the attacker extracts both the static model and the signatures, enabling attacks jointly optimized to evade both components (fourth row in \autoref{fig:tms}).
Again, the attacker selects an operating point corresponding to a 1\% FPR on an external dataset, without knowledge of the deployed thresholds.
This threat model is particularly relevant when dynamic analysis is performed in the cloud and static analysis is executed locally~\cite{windowsdefender_blog}.
By evading the initial stages, the attacker may gain sufficient time to compromise the system before dynamic analysis intervenes.

\mypar{Other Threat Models.}
In principle, enumerating all possible threat models would require considering every combination of assets an attacker might compromise, leading to an exponential number of cases.
In practice, this space is constrained by the availability of techniques and tools for evaluating the security of Compound AI Systems.
While a mature literature and stable open-source tools exist for adversarial attacks against static detectors~\cite{demetrio2021secml, song2022mab}, we are not aware of equally stable implementations targeting dynamic analysis (marked with the white symbol in \autoref{fig:concept}).
Proposed approaches either (i) lack public implementations~\cite{rosenberg2017bypassing}, (ii) rely on deprecated and unmaintained platforms such as the Cuckoo sandbox, discontinued in 2021~\cite{digregorio2024tarallo, rosenberg2018generic}, or (iii) operate only in \emph{feature space}~\cite{rosenberg2020query, chen2023malader}.
Therefore, also considering the observed superiority of static over dynamic analysis~\cite{dambra2023decoding}, we restrict our study to the four threat models in \autoref{fig:tms}, deferring the analysis of attacks against dynamic analysis components to future work.
\section{Implementation Details}
\label{sec:internals}
In this section, we present \textbf{\obelisk}, our implementation of a Compound AI System for malware detection, along with the attacks that follow the threat models described in \autoref{sec:methodology}.
The implementation of \obelisk and the attacks are publicly available at \textit{\url{https://github.com/Andrea-Ponte/obelisk}}.
We use this framework to quantify the trade-offs among performance, computational requirements, and robustness. 

\subsection{\obelisk Implementation}
\label{sec:obelisk_implementation}
\obelisk is implemented following the architecture shown in \autoref{fig:concept}, which is composed of three levels (see \autoref{sec:methodology}).

\mypar{Signature Matching with YARA.}
The first level is implemented using YARA rules\footnote{\url{https://github.com/VirusTotal/yara}}, a pattern-matching tool for signature-based detection.
Our system combines a blocklist, consisting of malicious signatures collected from public sources, with an allowlist, containing hashes of well-known programs distributed with Windows installations.
Although some rules achieve high detection rates, they also produce an unacceptable number of false alarms.
We hence discard all those that trigger at least one false alarm on the training data.
This choice reflects the fact that many rules were developed in isolation and not validated on large and heterogeneous datasets, making false alarms on benign software likely.

\mypar{Static Analysis with EMBER-GBDT.}
The second level is implemented using a Gradient Boosting Decision Tree (GBDT)~\cite{friedman2001greedy,friedman2002stochastic} from the XGBoost library~\cite{chen2015xgboost}, trained on features extracted with EMBER~\cite{anderson2018ember}.
Although this feature set was introduced in 2018, it remains one of the most adopted benchmarks in the literature~\cite{trizna2022quo,ponte2025slifer,ponte2025demystifying,kozak2025updating,yang2021cade,severi2021explanation}.

\mypar{Dynamic Analysis with Nebula.}
The third level is implemented using Nebula~\cite{trizna2024nebula}, a transformer-based model~\cite{vaswani2017attention} trained on reports generated with Speakeasy\footnote{\url{https://github.com/mandiant/speakeasy}\label{foo:speakeasy_url}}, a Python-based Windows kernel emulator.
We select Nebula because it provides an end-to-end detection approach that outperforms previous methods~\cite{jindal2019neurlux,zhang2020dynamic} on this type of data.

%\color{red}
\mypar{Error Management.} 
In this evaluation, we exclude samples that fail during preprocessing instead of assigning them a default label. This ensures that our assessment of Compound AI System trade-offs is not skewed by inflated false positive rates (from labeling errors as malicious) or degraded detection rates (labeling errors as benign), effect documented in~\cite{ponte2025slifer}.

\mypar{Training and Inference.}
We now describe how we implement the progressive reduction of the training and test data introduced in \autoref{sec:sys_form}.
Following the methodology outlined in \autoref{sec:methodology}, the static model at the second level is trained under two settings: first, on the full dataset, without considering the signature-based filtering level (\gbdtall); and second, on the subset of samples that are not identified by either the blocklist or the allowlist (\gbdtfilter).
We then use the probability scores produced by the resulting second-level static model to determine which samples should be used to train the computationally more expensive dynamic stage.
Similarly, the dynamic model is also trained under two settings: on the full dataset without considering the first and second levels (\nebulaall), and on the subset of samples that remain after applying the previous filtering levels (\nebulafilter).
In the latter case, we apply a two-step filtering procedure.
First, we retain only the samples that are not identified by either the blocklist or the allowlist.
Then, we introduce a parameter $\bm{\delta}$ that defines two decision thresholds on the static model's estimated probability that a sample is malware.
Training samples with probability $\bm{p \leq \delta}$ are labeled as benign, while samples with probability $\bm{p \geq 1-\delta}$ are labeled as malicious.
Consequently, only the samples whose probability lies in the interval $\bm{(\delta, 1-\delta)}$, which we interpret as the uncertainty region of the static model, are used to train the dynamic model.
Such a decision contrasts with the standard approach, where detectors are only using a single threshold to determine the maliciousness of a sample (i.e., a sample is considered malicious when the computed probability exceeds a certain threshold).
In fact, with only one detection threshold, it is not possible to reduce the computational requirements of a Compound AI System, as all benign programs would be sent to the costly dynamic analysis level~\cite{ponte2025slifer}.
At inference time, the blocklist and allowlist are first used to label known samples as malicious or benign, respectively.
Samples that do not match any signature are then passed to the static model, which assigns each of them a probability $p$ of being malware.
Those with $\bm{p \leq \delta}$ are classified as benign, whereas those with $\bm{p \geq 1-\delta}$ are classified as malicious.
The remaining samples, i.e., those falling in the static model's uncertainty region, are forwarded to the dynamic stage for the final decision.

\subsection{Attacks Implementation}
\label{sec:tm_implementation}
To evaluate the adversarial robustness of \obelisk, we rely on two established techniques: \emph{GAMMA section injection}~\cite{demetrio2021functionality} (GAMMA), which inserts non-executable content extracted from benign programs, and \emph{padding} attacks~\cite{kolosnjaji2018adversarial} (PAD), which append bytes to the overlay.
Both techniques are used to implement attacks for each threat model introduced in \autoref{sec:robustness} and illustrated in \autoref{fig:tms}.

\mypar{Attacks for TM1 (A1).}
We construct a surrogate for \obelisk following a standard approach: we leverage a pre-trained open-source model\footnote{\url{https://github.com/endgameinc/malware_evasion_competition/tree/master/models/ember}\label{foo:surrogateA1}}, which in our case is a GBDT classifier trained on the EMBER dataset~\cite{anderson2018ember}.
We then generate two sets of adversarial examples against this model using the GAMMA and PAD attacks.

\mypar{Attacks for TM2 (A2).}
In TM2, we have exact knowledge of the first layer of the deployed Compound AI System, namely the signatures.
A2 is therefore constructed by exploiting this information from A1.
Since YARA rules are triggered by specific artifacts, we enhance GAMMA by explicitly avoiding the injection of artifacts that could activate such signatures, while still inserting carefully crafted content to evade detection; we denote this variant as GAMMA-YARA.
We then employ GAMMA-YARA and PAD to attack a surrogate Compound AI System consisting of the signatures followed by the model adopted in A1$^{\text{\ref{foo:surrogateA1}}}$.
We thus generate two sets of adversarial examples against this model, one using PAD and the other using GAMMA-YARA.

\mypar{Attacks for TM3 (A3).}
A3 is identical to A1, except that, rather than relying on the surrogate used in A1$^{\text{\ref{foo:surrogateA1}}}$, we use the same GBDT model deployed in \obelisk.
Note that \obelisk may employ different GBDT models depending on the training set considered (see \autoref{sec:obelisk_implementation}).
Accordingly, we generate different sets of adversarial examples against this model, depending on both the attack employed (PAD or GAMMA) and the GBDT variant considered: either the model trained on the full dataset (\gbdtall) or the one trained on the data filtered according to the YARA rules (\gbdtfilter).

\mypar{Attacks for TM4 (A4).}
A4 is identical to A2, except that, as A3 relates to A1, instead of relying on a surrogate model, we use the same GBDT model deployed in \obelisk.
Also in this case, \obelisk may employ different GBDT models depending on the training set considered (see \autoref{sec:obelisk_implementation}).
We therefore generate different sets of adversarial examples against this model, depending on both the attack employed (PAD or GAMMA-YARA) and the GBDT variant considered (\gbdtall and \gbdtfilter).
\section{Experimental Analysis}
\label{sec:experiments}
In this section we first detail our experimental setup (\autoref{sec:setup}) and then report our findings in terms of trade-off between performance and computational requirements (\autoref{sec:performance}), robustness and computational requirements (\autoref{sec:adv_evaluation}), along with a dedicated analysis of the detection levels of the Compound AI Systems (\autoref{sec:levels}). We summarize findings and analysis in take-home messages (\textbf{THM}) and practical deployment guidelines (\textbf{G}).
\subsection{Experimental Setup}
\label{sec:setup}
\mypar{Dataset.}
We use the Speakeasy dataset~\cite{trizna2022quo} as the main data source.
It contains PE files divided into training and test splits, collected in January 2022 and April 2022, respectively, and includes malware samples from seven families: Backdoor, Coinminer, Dropper, Keylogger, Ransomware, RAT, and Trojan.
After duplicate removal, the training set consists of $71{,}505$ malware samples and $26{,}059$ benign samples, while the test set contains $17{,}495$ malware samples and $10{,}000$ benign samples.
To mitigate the class imbalance, we increase the number of benign samples by adding $2{,}648$ Windows system files from the \textit{sys32} and \textit{syswow64} directories of Windows 8.1, 10, and 11, together with $9{,}809$ programs obtained from Chocolatey\footnote{\url{https://chocolatey.org/}\label{foo:chocolatey}}.
To avoid data snooping~\cite{arp2022and}, Windows system files are included only in the training set, whereas Chocolatey samples, collected in 2024, are reserved for testing.
The final PE dataset therefore comprises $71{,}505$ malware samples and $28{,}707$ goodware samples for training, and $17{,}495$ malware samples and $19{,}809$ goodware samples for testing.

\mypar{Signatures.}
We collected the signatures (\autoref{sec:obelisk_implementation}) forming the blocklist in April 2025 from several open-source repositories\footnote{\url{https://github.com/bartblaze/Yara-rules/tree/master/rules}}$^,$\footnote{\url{https://github.com/elastic/protections-artifacts/tree/main/yara/rules}}$^,$\footnote{\url{https://github.com/malpedia/signator-rules/tree/main/rules}}$^,$\footnote{\url{https://github.com/Neo23x0/signature-base/tree/master/yara}}$^,$\footnote{\url{https://github.com/Yara-Rules/rules/tree/master/malware}}$^,$\footnote{\url{https://yaraify.abuse.ch/yarahub/}}, excluding those that produced any false positive on the training data, yielding $236$ signatures.
We also collected $2{,}648$ hashes from Windows 8.1, 10, and 11 system files to build the allowlist.
The blocklist and the allowlist are integrated into \obelisk using YARA Python\footnote{\url{https://github.com/VirusTotal/yara-python}}.
%

\begin{comment}

\begin{table}[]

\centering
\begin{tabular}{@{}ccccc@{}}
\toprule
\multirow{2}{*}{\textbf{Ruleset}} & \textbf{True} & \textbf{False} & \textbf{Detection} & \textbf{False Positive} \\
& \textbf{Positive} & \textbf{Positive} & \textbf{Rate} & \textbf{Rate} \\ \midrule
\textbf{Original} & 17385 & 17514 & 0.9937 & 0.884 \\ 
\textbf{Cleansed} & 4244 & 115 & 0.2426 & \textbf{0.0058} \\
\bottomrule
\end{tabular}
\caption{Performance of the rules on the test set, before and after filtering to achieve a $0\%$ false positive rate on the training dataset.
We report both the absolute number of detected samples and false alarms, and their corresponding ratios with respect to the entire test set.}
\label{tab:rules_on_testset}
\end{table} 

\end{comment}
%

\mypar{Static Analysis using GBDT.}
The GBDT model used in the static level (see \autoref{sec:obelisk_implementation}) is trained 
%on EMBER features~\cite{anderson2018ember}, 
with $1000$ trees, a learning rate of $0.1$, and $32$ parallel jobs, while keeping the other parameters to their default values~\cite{chen2015xgboost}.

\mypar{Dynamic Analysis using Nebula.}
This phase relies on behavioral reports generated with the Speakeasy emulator$^\text{\ref{foo:speakeasy_url}}$.
We use both the reports released with the dataset~\cite{trizna2022quo} and additional reports generated for the newly collected goodware samples from Chocolatey$^\text{\ref{foo:chocolatey}}$.
As observed in previous work~\cite{trizna2022quo, ponte2025slifer}, some files fail during emulation, thereby reducing the number of usable reports.
The dataset also contains duplicated reports originating from different samples, i.e., samples identified by different SHA-256 hashes.
Upon inspection, we found that some of these correspond to malware variants, whereas others arise from limitations of the emulator.
We remove all errors from the training set, counting $78{,}084$ successful reports.
Then we retain unique reports for a total of $67{,} 611$.
Lastly, at test time, we evaluated all systems discarding samples that could not be analyzed as anticipated in \autoref{sec:obelisk_implementation}.
The Nebula model used at the dynamic level (see \autoref{sec:obelisk_implementation}) is built using the same hyperparameters and preprocessing steps (filtering, normalization, and tokenization) reported by the authors in the original paper~\cite{trizna2024nebula} and the associated library\footnote{\url{https://github.com/dtrizna/nebula}}.
Specifically, the tokenizer is implemented using BPE~\cite{gage1994new} with a vocabulary of $50{,}000$ tokens, and the model takes as input sequences of up to $512$ tokens.
We train the model for $60$ epochs using the AdamW~\cite{loshchilovdecoupled} optimizer, with a learning rate of $\alpha=1 \times 10^{-4}$, $\beta_1 = 0.9$, $\beta_2 = 0.999$, and $\epsilon = 10^{-8}$ and a batch size of $64$.
The best model is selected based on the validation ROC AUC computed on a $10\%$ split of the training data.

\begin{comment}

\begin{table}
\centering
\begin{tabular}{@{}cccc@{}}
\toprule
\textbf{Dataset} & \textbf{Successful} & \textbf{Errors} & \textbf{Error Rate}\\ \midrule
\textbf{Train} & 78084 & 22128 & 0.2208 \\
\textbf{Test} & 24211 & 13093 & 0.3510\\ \bottomrule
\end{tabular}
\caption{Emulation reports and errors for the training and test sets, together with the corresponding error rates.}
\label{tab:emulation_errors}
\end{table}

\end{comment}

\mypar{Setup of Adversarial Attacks.}
For A1--A4, we generated the adversarial sample sets described in \autoref{sec:tm_implementation} using GAMMA, GAMMA-YARA, and PAD.
For GAMMA, we leveraged the injection of $50$ \texttt{.rdata} sections from Chocolatey samples (GAMMA-CHOCO) and $10$ sections from Windows 11 files (GAMMA-WIN), while setting the regularization parameter to $\lambda = 10^{-7}$.
Note that GAMMA-YARA is only needed when injecting sections from Chocolatey, as these may sometimes trigger signatures~\cite{ponte2025demystifying, ponte2025slifer}, yielding the GAMMA-YARA-CHOCO variant.
Conversely, GAMMA-YARA-WIN is not reported, since sections from Windows 11 files never trigger signatures and, therefore, GAMMA-WIN coincides with GAMMA-YARA-WIN.
For PAD, we configured attacks that inject $0.5$\,KB, $1$\,KB, and $1.5$\,KB of content.
Because all the described attacks share the same genetic algorithm as optimizer~\cite{demetrio2021functionality, fortin2012deap}, we set a maximum of $500$ queries and a population size of $10$, i.e., the number of candidate solutions evaluated at each optimization round.
All attacks were instantiated through the SecML Malware library~\cite{demetrio2021secml}. We generated all attacks using the same pool of 700 malware samples drawn from the Speakeasy testset, comprising 100 samples from each family.
Thus, we consider the following sets of adversarial EXEmples for each attack:
\begin{itemize}
\item A1 comprises three sets of adversarial samples, generated with GAMMA-CHOCO, GAMMA-WIN, and PAD against a surrogate model;
\item A2 comprises three sets of adversarial samples, generated with GAMMA-YARA-CHOCO, GAMMA-WIN, and PAD.
These attacks are in part the same as A1 (i.e., GAMMA-WIN and PAD) but against a different model (YARA plus the surrogate model of A1);
\item A3 comprises six sets of adversarial samples, generated with GAMMA-CHOCO, GAMMA-WIN, and PAD against GBDT-ALL or GBDT-YARA;
\item A4 comprises six sets of adversarial samples, generated with GAMMA-YARA-CHOCO, GAMMA-WIN, and PAD against GBDT-ALL or GBDT-YARA.
These attacks are in part the same as A3 (i.e., GAMMA-WIN and PAD) but against a different model (YARA plus GBDT-ALL or YARA plus GBDT-YARA).
\end{itemize}

\mypar{Considered Compound AI Systems.}
We tested all the adversarial samples previously described against different variants of our Compound AI System \obelisk, named as follows:
\begin{itemize}
\item Standard (\standard), where the first level is implemented with YARA rules, the second level with \gbdtall with only one detection threshold, and the third level is implemented with Nebula trained on all reports (\nebulaall);
%(i.e., the de facto standard of malware detectors)

\item \obvone, which uses the same structure of \standard, but, at inference time, the second level filters data using two thresholds (YARA, \gbdtall, \nebulaall);
\item \obvtwo, which uses the same structure of \obvone, but the second level deploys \gbdtfilter (hence YARA, \gbdtfilter, \nebulaall);
\item \obvthree, which uses the same structure of \obvtwo but, the third level is implemented with Nebula trained on the data filtered by both YARA and \gbdtfilter (\nebulafilter), resulting in YARA, \gbdtfilter, \nebulafilter.
\end{itemize}

\mypar{Metrics.}
We quantify the detection performance with three different metrics: the True Positive Rate (TPR), counting the fraction of correctly classified malicious samples; the False Positive Rate (FPR), counting the fraction of raised false alarms, and the F1 Score (F1), which jointly accounts for both TPR and FPR.
We also include the transfer rate (TR), which counts the fraction of attacks that evade both the target for which they are computed and the real target system.
We quantify the computational requirements with two other metrics: the training time ($T_t$) and mean inference time ($T_i$), both expressed in seconds.
Training time ($T_t$) is computed as the sum of the average preprocessing time needed to extract EMBER features, the time needed to train GBDT, the average preprocessing time to produce reports with Speakeasy along with filtering and normalization, the training of the BPE tokenizer, and the time needed to train Nebula.
%TODO: LucaD unire train di Nebula e tokenizer
We omit from this analysis the time needed to produce signatures, as it would require estimating the amount of working days needed by reverse engineers (thus being infeasible for the scope of this paper).
Inference time $T_i$ is computed as the average time needed to make a prediction using the different versions of \obelisk. It sums the time needed to perform the pattern-matching with signatures (always) and possibly (if the signatures do not label it) the time to extract EMBER features and compute a probability with GBDT and possibly (if the prediction does not reach the desired confidence) the time needed to produce a report with Speakeasy, filtering and normalizing it, and computing a probability with Nebula.

\mypar{Hardware.}
All experiments have been conducted on a workstation equipped with an Intel® Xeon(R) Gold 5420, two Nvidia L40S GPUs, and 540 GB of RAM.

\begin{figure*}[tbp]
    \centering
    
    \begin{subfigure}[b]{0.3\textwidth}
        \centering
        \includegraphics[width=\textwidth]{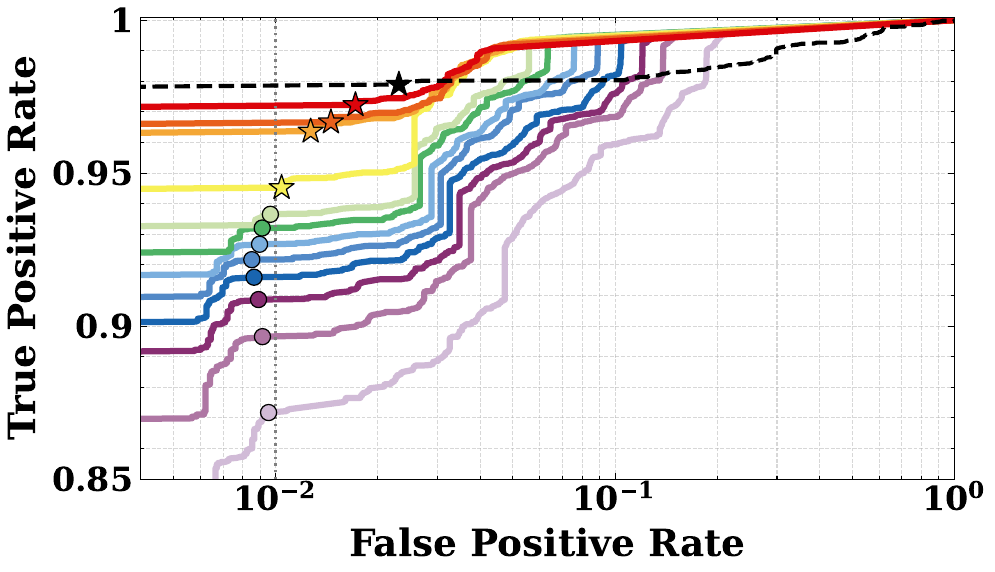}
        \caption{\obvone}
        \label{fig:roc_obv1}
    \end{subfigure}
    \begin{subfigure}[b]{0.3\textwidth}
        \centering
        \includegraphics[width=\textwidth]{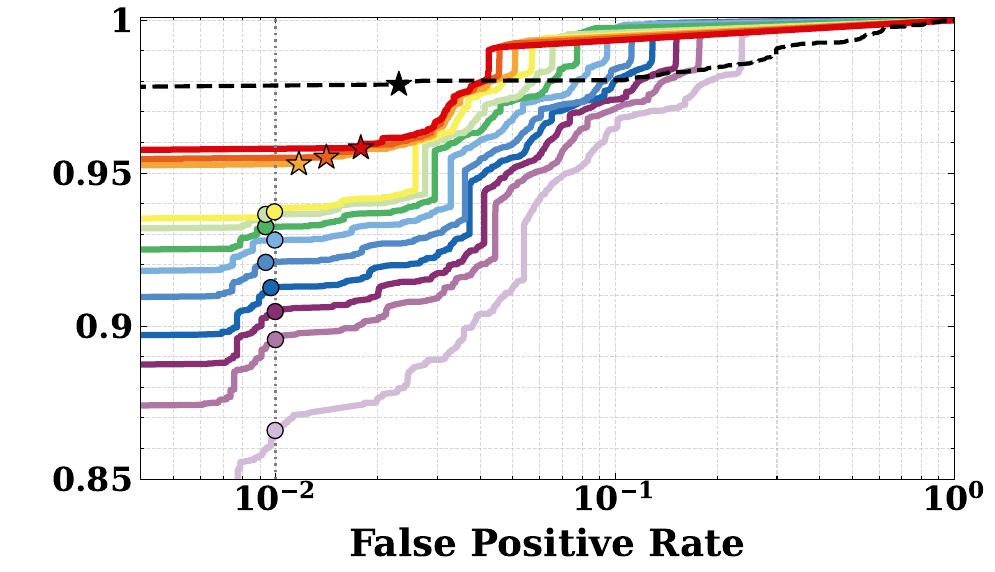}
        \caption{\obvtwo}
        \label{fig:roc_obv2}
    \end{subfigure}
    \begin{subfigure}[b]{0.3\textwidth}
        \centering
        \includegraphics[width=\textwidth]{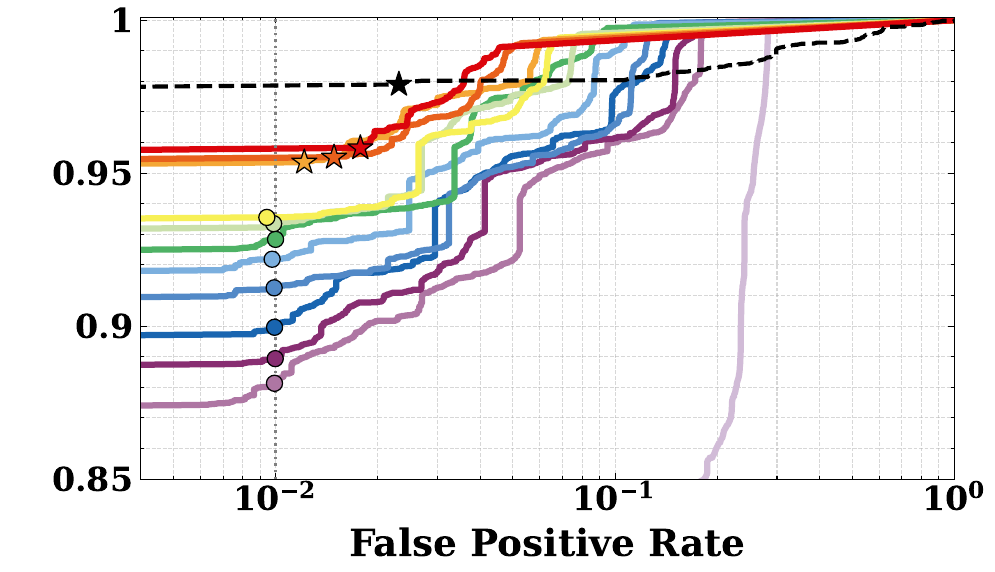}
        \caption{\obvthree}
        \label{fig:roc_obv3}
    \end{subfigure}

    \includegraphics[width=0.9\textwidth]{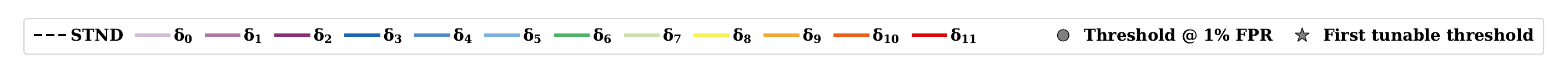}
    
    \caption{Receiver Operating Characteristic (ROC) curves for the evaluated Compound AI systems. From (a) to (c), we plot the different configurations of \obvone, \obvtwo and \obvthree, keeping the \standard system always depicted as a black dashed line. }
    \label{fig:roc_curves_all}
\end{figure*}

\begin{figure*}[tbp]
    \centering
    
    \begin{subfigure}{0.49\textwidth}
        \centering
        \includegraphics[width=\textwidth]{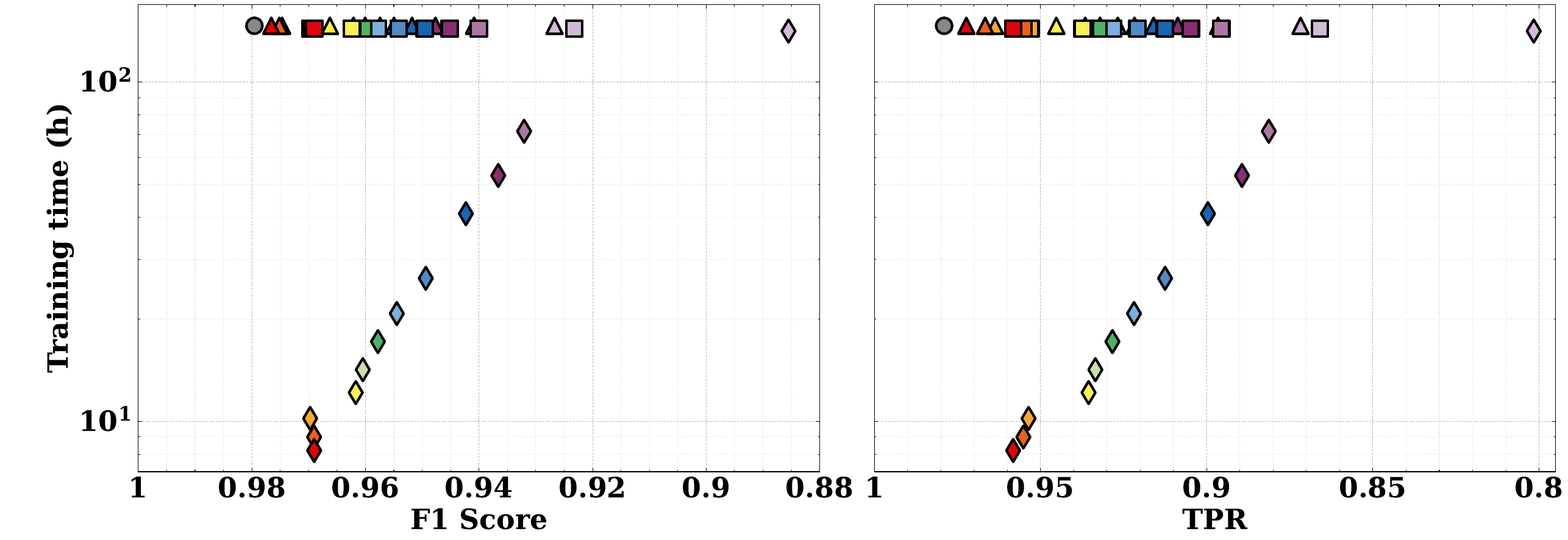}
        \caption{Training Time vs Performance}
        \label{fig:pareto_perf_training}
    \end{subfigure}
    \hfill
    \begin{subfigure}{0.49\textwidth}
        \centering
        \includegraphics[width=\textwidth]{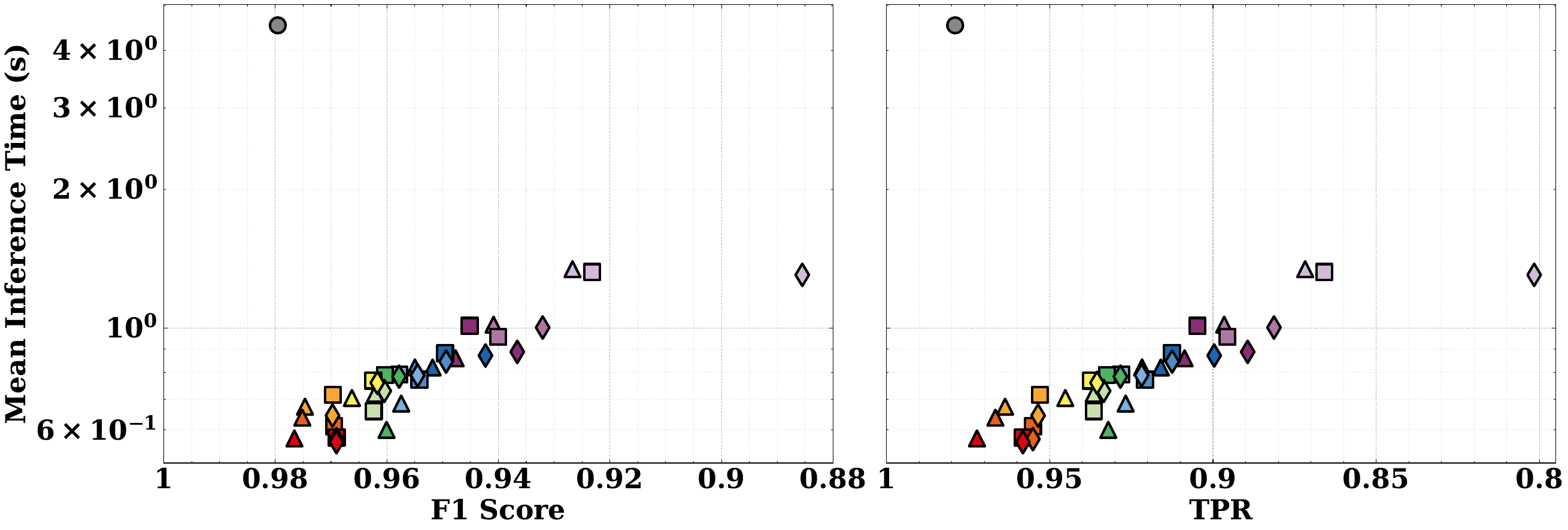}
        \caption{Mean Inference Time vs Performance}
        \label{fig:pareto_perf_inference}
    \end{subfigure}
    
    \centering
    \includegraphics[width=0.9\textwidth]{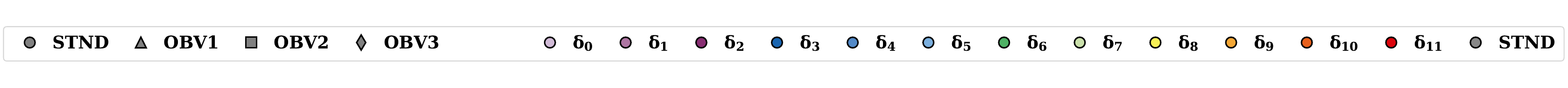}
    
    \caption{Pareto fronts representing the trade-off between performance and $T_t$ (a) and between performance and $T_i$ (b). We considered all 37 Compound AI Systems using different markers and colors. }
    \label{fig:pareto_performance}
\end{figure*}
 
\subsection{Performance vs Computational Requirements}
\label{sec:performance}
To quantify the trade-off between detection performance and computational requirements, we first conduct an ablation study on the threshold $\delta$ that \obvone, \obvtwo, and \obvthree use to filter samples between levels.
We consider twelve increasing values $(0.1, 0.27, 0.44, 0.72, 1.18, 1.93, 3.16, 5.18, 8.48, 13.89, 22.76,\allowbreak 37.28) \times 10^{-3}$, denoted in ascending order as $\delta_0, \delta_1, \ldots, \delta_{11}$.
This yields twelve configurations for each of \obvone, \obvtwo, and \obvthree which, together with the \standard baseline, amount to the 37 Compound AI Systems considered throughout our evaluation.
The amount of data reaching each level differs across systems, and so does the cost of training it.
\gbdtall, deployed in \standard and \obvone, is trained on the full training set of $100{,}212$ samples, whereas \gbdtfilter, deployed in \obvtwo and \obvthree, is trained only on the $79{,}319$ samples that are not resolved by the signature level.
Likewise, \nebulaall, deployed in \standard, \obvone, \obvtwo, and \obvthree, is trained on all $67{,}611$ available emulation reports, while \nebulafilter, deployed in \obvthree, is trained only on the reports that fall inside the uncertainty region of the static level; its training set shrinks from $59{,}461$ samples for $\delta_0$ down to $3{,}373$ for $\delta_{11}$.
We evaluate all systems on the test set, reporting the resulting ROC curves in \autoref{fig:roc_curves_all}, namely \obvone in \autoref{fig:roc_obv1}, \obvtwo in \autoref{fig:roc_obv2}, and \obvthree in \autoref{fig:roc_obv3}, each compared against \standard (black dashed line) and each solid curve corresponding to one value of $\delta$.
These curves expose a consistent effect of $\delta$ on test-time performance: the TPR degrades when a large fraction of samples is forwarded to the dynamic level (i.e., low values of $\delta$), and improves as filtering becomes more aggressive (i.e., high values of $\delta$).
This is expected, since delegating more decisions to the dynamic level exposes the system to its lower stand-alone accuracy~\cite{dambra2023decoding}.
To obtain scores that are comparable across systems, we tune the detection threshold of each system (i.e., the threshold of its last level) at ${\approx}1$\% FPR, selecting the point on the ROC curve that comes closest to this constraint and therefore attaining slightly smaller or slightly higher FPRs (circles and stars in \autoref{fig:roc_curves_all}, respectively).
On this common operating point we can inspect the trade-off between performance and computational requirements in \autoref{fig:pareto_performance}, which relates TPR and F1 to $T_t$ in \autoref{fig:pareto_perf_training}, and to $T_i$ in \autoref{fig:pareto_perf_inference}, for each of the 37 Compound AI Systems.
When considering the trade-off between $T_t$ and performance:
\begin{enumerate}[label=\textbf{THM\arabic*}, start=1, leftmargin=1.21cm,style=nextline,itemindent=.0cm]
    \item \obvthree is the fastest family of systems to train while retaining high detection performance, the clearest example being \obvthree with $\delta_{11}$, which scores ${\approx}96$\% TPR at ${\approx}1$\% FPR.\label{THM1}
    \item \standard, \obvone, and \obvtwo require an extensive training time ($\geq 100$ hours), dominated by data preparation rather than by model fitting: feature extraction costs ${\approx}0.26$s per sample and, above all, emulation requires ${\approx}8.1$s per sample.\label{THM2}
\end{enumerate}
\noindent
When observing the trade-off between $T_i$ and performance:
\begin{enumerate}[label=\textbf{THM\arabic*}, start=3, leftmargin=1.21cm,style=nextline,itemindent=.0cm]
    \item The best system is \obvone with $\delta_{11}$, but almost all the others cluster in the bottom-left corner of the plot, with \obvthree at $\delta_{11}$ nearly matching the performance of the best system at a comparable inference cost.\label{THM3}
    \item \standard requires $\geq 4$ seconds per sample on average, being one order of magnitude slower than its competitors.
    This delay follows directly from its design: malware samples are typically stopped by an early level, whereas goodware samples traverse the entire pipeline and therefore always incur both feature extraction and emulation.
    Since inference is by far the most frequent operation, such a delay is likely to be unacceptable in a production environment.\label{THM4}
\end{enumerate}
\noindent
These results indicate that \obvthree with $\delta_{11}$ is the system that best balances performance and computational cost, making it suitable for production environments as long as only these two aspects are considered.

\begin{figure*}[tbp]
    \centering
    
    \begin{subfigure}{\textwidth}
        \centering
        \includegraphics[width=\textwidth]{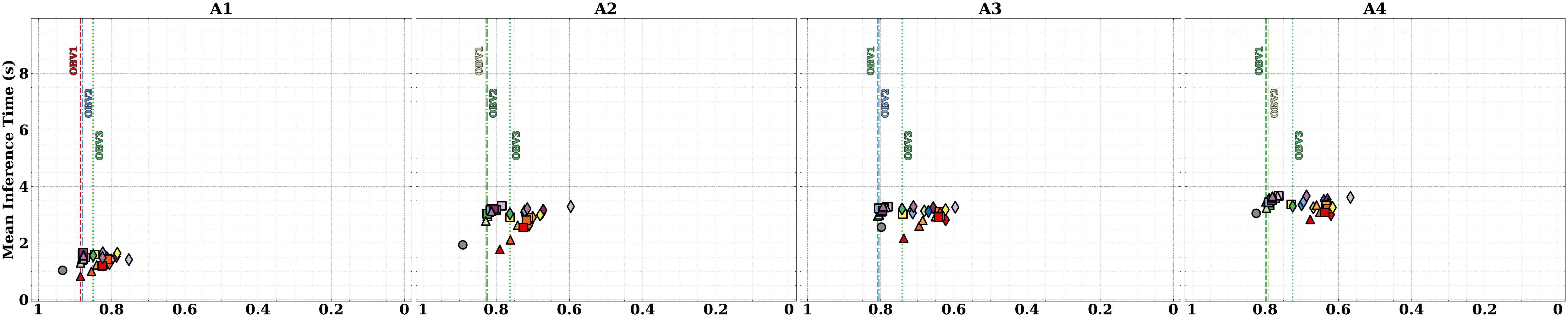}
        \caption{Trade-off between Worst-case TPR and $T_i$ computed on all 37 Compound AI Systems for all attacks A1--A4.}
        \label{fig:pareto_rob_tpr}
    \end{subfigure}
        
    \begin{subfigure}{\textwidth}
        \centering
        \includegraphics[width=\textwidth]{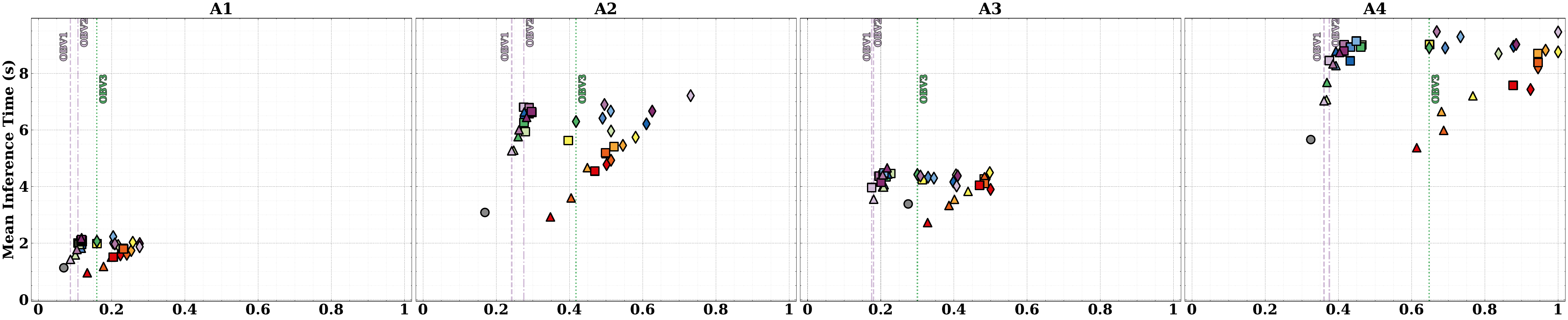}
        \caption{Trade-off between Worst-case TR and $T_i$ computed on all 37 Compound AI Systems for all attacks A1--A4.}
        \label{fig:pareto_rob_transfer}
    \end{subfigure}
    
    \includegraphics[width=0.9\textwidth]{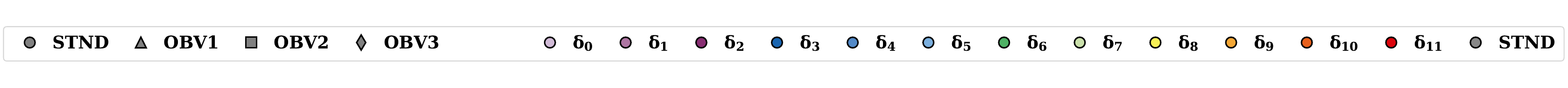}
    
    \caption{Pareto fronts representing the trade-off between $T_i$ and worst-case TPR (a) and TR (b). We considered all 37 Compound AI Systems using different markers and colors. }
    \label{fig:pareto_robustness}
\end{figure*}

\subsection{Adversarial Robustness vs Computational Requirements}
\label{sec:adv_evaluation}
We now turn to the trade-off between robustness and computational requirements when the systems are under attack.
We take the 37 Compound AI Systems tuned at ${\approx}1$\% FPR and, against each of them, we evaluate the adversarial EXEmples generated under the four threat models, i.e., A1--A4 (see \autoref{sec:tm_implementation}).
We quantify robustness through the \emph{worst-case} TPR, computed by treating the attacks available within each threat model as an ensemble~\cite{cina2025attackbench, croce2robustbench}.
For every sample, the attack is counted as successful if at least one strategy in the ensemble yields an adversarial EXEmple that evades the target system.
The TPR computed on these outcomes therefore describes the worst-case scenario in which the attacker can always select the most effective strategy on a per-sample basis, and it is the metric we report throughout this section.
\autoref{fig:pareto_robustness} shows the resulting trade-off for all 37 systems: the interplay between $T_i$ and the worst-case TPR in \autoref{fig:pareto_rob_tpr}, and the interplay between $T_i$ and TR in \autoref{fig:pareto_rob_transfer}.
From these results we report that:
\begin{enumerate}[label=\textbf{THM\arabic*}, start=5, leftmargin=1.21cm,style=nextline,itemindent=.0cm]
    \item \standard is the most robust system across almost all threat models, and it ties with the other top performers under A3.
    It is followed by \obvone and \obvtwo (at intermediate values of $\delta$), while \obvthree is the least robust family of systems.
    Neither training the levels on less data nor filtering samples at the static level is therefore beneficial for robustness.\label{THM5}
    \item Robustness is not monotone in $\delta$: both \obvone and \obvtwo attain their best robustness at intermediate values ($\delta_6$ and $\delta_7$), and the most robust configuration of \obvthree is again $\delta_6$, at the cost of an average TPR drop of less than $10\%$ w.r.t. \standard.\label{THM6}
    \item The TR grows as attacker knowledge increases from TM1 to TM4, since knowing more levels of the target lets the attacker craft attacks that are tailored to the deployed components rather than to a surrogate.\label{THM7}
    \item Attacks inflate $T_i$ since (i) the manipulations enlarge the input, slowing down feature extraction at the static level, and  (ii) the adversarial EXEmples that bypass the first two levels reach the dynamic level, triggering emulation, i.e., the most expensive operation of the pipeline.
    Thus, robustness and responsiveness are coupled,
    as attacks degrade the latter even when they do not fully defeat the former.\label{THM8}
\end{enumerate}
\noindent
Summarizing, \standard is the best choice for the trade-off between robustness and computational requirements under attack: it does not rely on a filtering threshold that widens the attack surface, and its levels are trained on more data, which makes them better at flagging samples that were never observed at training time.

\subsection{Level-wise Analysis}
\label{sec:levels}
\begin{figure}
    \centering
    \includegraphics[width=0.88\linewidth]{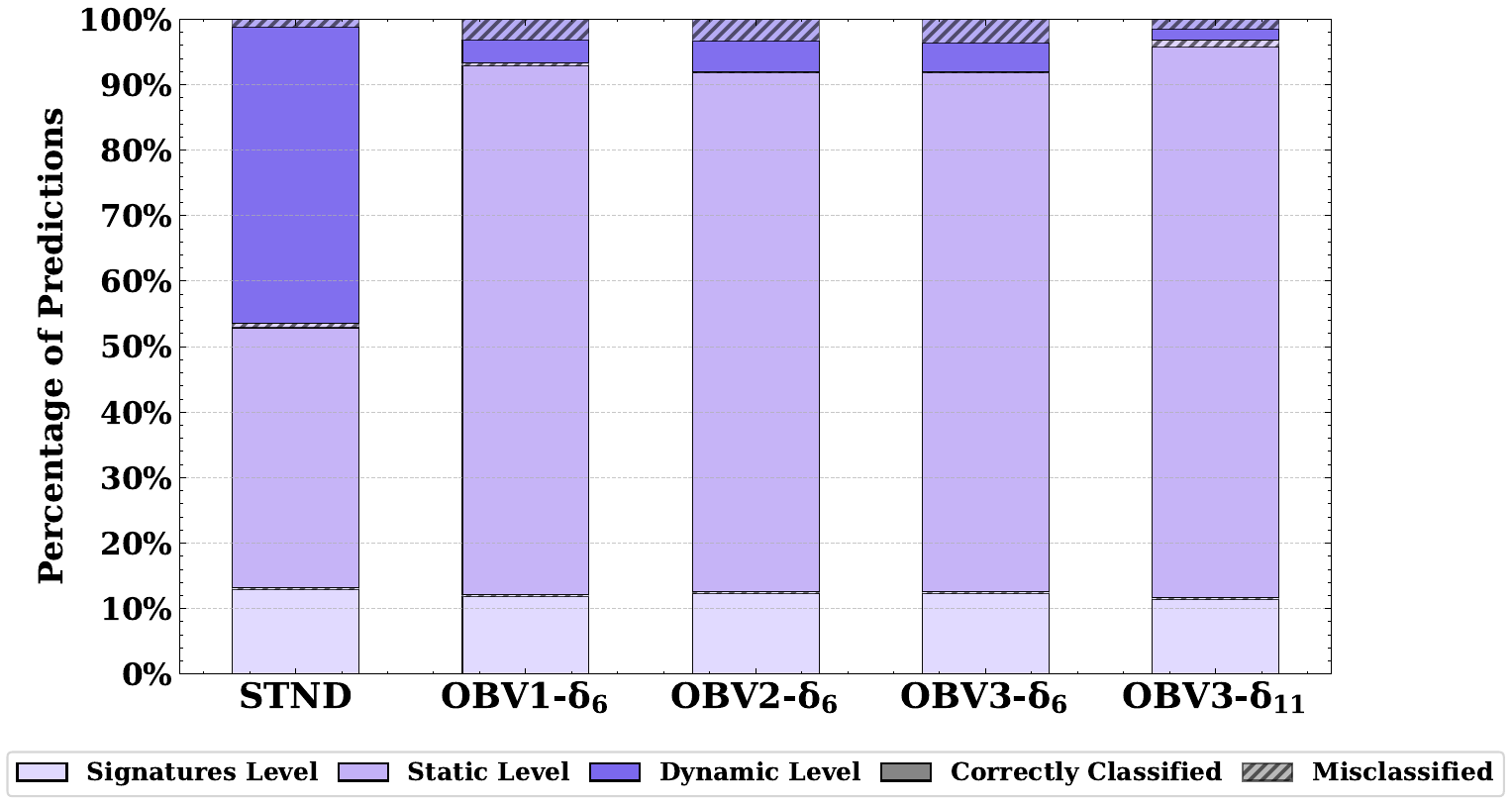}
    \caption{Percentage of test-set predictions per level, with correct (solid) and misclassified (hatched) samples shown separately.}
    \label{fig:levels_performance}
\end{figure}

\begin{figure}[t]
    \centering
    \includegraphics[width=0.95\linewidth]{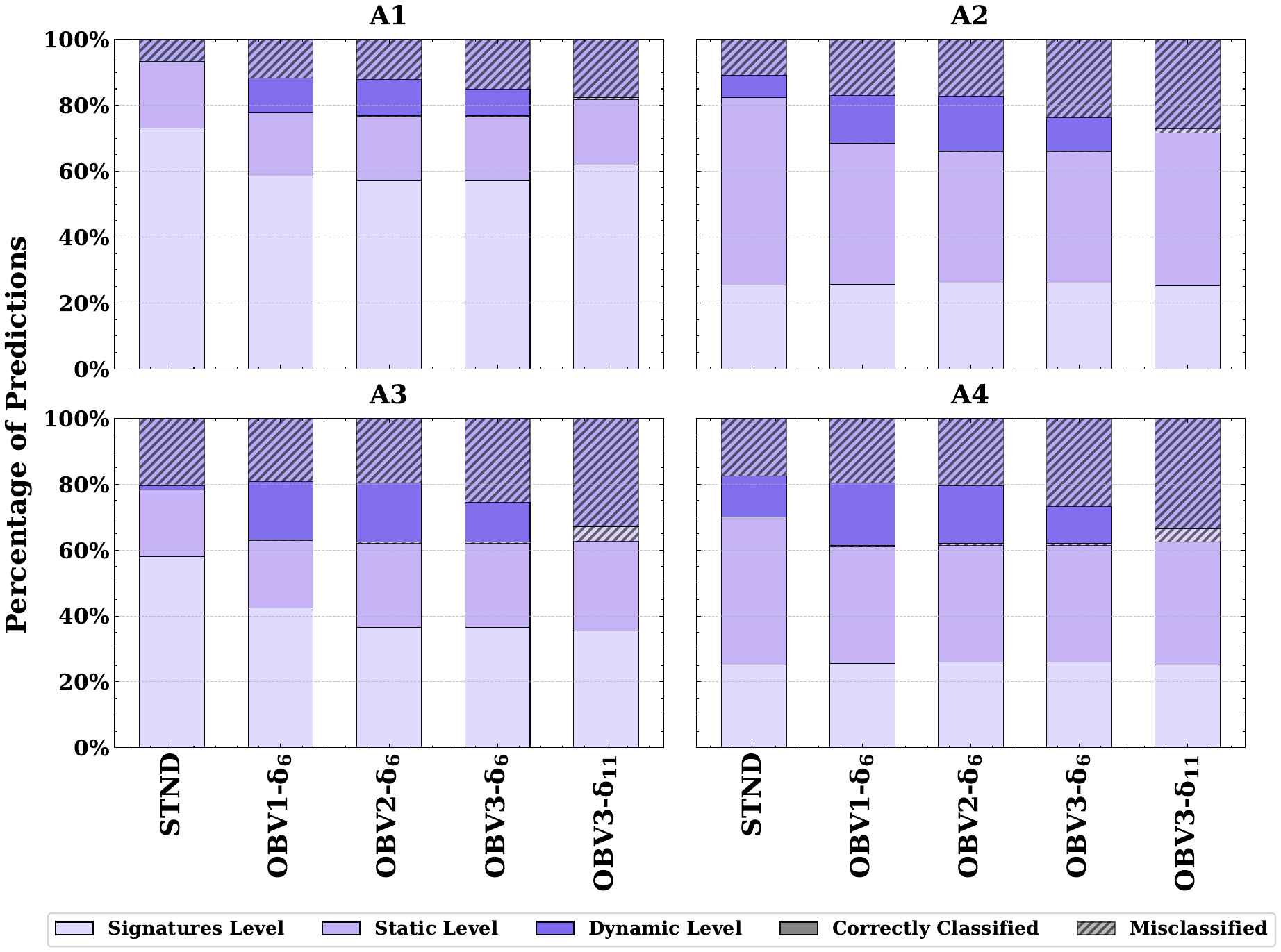}
    \caption{Percentage of predictions under attacks A1–A4 per level, with correct (solid) and misclassified (hatched) samples shown separately.}
    \label{fig:levels_robustness}
\end{figure}

We now analyze how each level of the Compound AI Systems contributes to both performance and robustness.
Rather than repeating this analysis on all 37 systems, we focus on five representative ones:
\standard, as the most accurate and robust system, at the expense of computational requirements (\ref{THM5});
\obvone and \obvtwo with $\delta_6$, as they exhibit average behavior across all metrics (\ref{THM3} and \ref{THM6});
\obvthree with $\delta_6$, being among the fastest systems to train while incurring the smallest robustness drop within its family (\ref{THM1} and \ref{THM6});
and \obvthree with $\delta_{11}$, the fastest system to train and among the fastest at inference time, retaining top-tier performance at the expense of robustness (\ref{THM1} and \ref{THM3}).
For each of these systems, \autoref{fig:levels_performance} reports the percentage of test samples whose final decision is taken by a specific level, separating correct predictions from misclassifications (solid and hatched bars, respectively).
This analysis shows that:
\begin{enumerate}[label=\textbf{THM\arabic*},
start=9,
labelsep=0.15cm,
leftmargin=*,
align=left,
widest=10
]
    \item The filtering threshold $\delta$ shifts the decision burden onto the static level, which resolves ${\approx}80\%$ of the test set on its own.
    \standard behaves in the opposite way, delegating most benign samples to the dynamic level, which accounts for ${\approx}47\%$ of the correctly classified test set and explains the $T_i$ reported in \ref{THM4}.\label{THM9}
    \item The signature level correctly classifies ${\approx}13\%$ of the test set while raising a negligible number of false alarms, a direct consequence of discarding every signature that fires false positives on the training set.\label{THM10}
\end{enumerate}
\noindent
Turning to robustness, \autoref{fig:levels_robustness} reports the same level-wise breakdown when the systems are exposed to the adversarial EXEmples of A1--A4, under the worst-case scenario described in \autoref{sec:adv_evaluation}.
Here we observe that:
\begin{enumerate}[label=\textbf{THM\arabic*}, start=11,
labelsep=0.15cm,
leftmargin=*,
align=left,
widest=10]
    \item Signatures are effective defenders, but only if the deployed set stays undisclosed: for \standard, the share of attacks stopped at this level falls from ${\approx}75\%$ under A1 and A3 to ${\approx}25\%$ under A2 and A4, where the attacker avoids the artifacts that trigger the rules.\label{THM11}
    \item The filtering threshold $\delta$ should not be set too high, as an overly wide benign region lets the attacker defeat the entire pipeline by evading the static level alone; this is visible as the hatched area on top of the static level for \obvthree with $\delta_{11}$.\label{THM12}
    \item The dynamic level contributes to defense only when trained on the full dataset. The model deployed in \obvthree with $\delta_{11}$, trained on the smallest number of reports, stops virtually no attack once the first two levels have been bypassed.\label{THM13}
\end{enumerate}
\noindent
Taken together, \ref{THM11}--\ref{THM13} show that the robustness of a Compound AI System is not the sum of the robustness of its levels: aggressive filtering both removes decisions from the levels best placed to catch unknown samples and hands the attacker a shortcut through the pipeline.


\subsection{Final Recommendations}\label{sec:recommendations}
We conclude by distilling our findings into four deployment guidelines, each tailored to a distinct set of operational constraints, so that practitioners can identify the configuration matching their own setting:
\begin{enumerate}[label=\textbf{G\arabic*}, start=1, leftmargin=*, style=nextline, itemindent=0cm]
\item \textbf{No dominant constraint.}
Deploy \obvthree with $\delta_6$, i.e., a system whose levels are all trained on filtered data and which also filters samples at inference time.
It is the balanced compromise across all metrics, excelling at none of them but never collapsing on any (\ref{THM1}, \ref{THM6}).\label{G1}
\item \textbf{Computation constrained at both training and inference time.}
Deploy \obvthree with $\delta_{11}$, the fastest system to train and among the fastest to compute predictions, while still reaching ${\approx}96\%$ TPR at ${\approx}1\%$ FPR (\ref{THM1}, \ref{THM3}).
This speed is paid in robustness: its wide benign region lets attackers bypass the pipeline by evading the static level alone, and its dynamic level, trained on the fewest reports, provides no residual defense (\ref{THM12}, \ref{THM13}).\label{G2}
\item \textbf{Training resources abundant, deployment resources constrained.}
Deploy \obvone with $\delta_6$, which combines strong performance, robustness close to \standard, and fast inference, at the price of training cost (\ref{THM2}, \ref{THM6}).
\item \textbf{Performance and robustness paramount, latency not a bottleneck.}
Deploy \standard, the most accurate and most robust system in our evaluation (\ref{THM5}).
It is viable only where the infrastructure tolerates decisions taken in the order of seconds per sample, as every benign program traverses the full pipeline (\ref{THM4}, \ref{THM9}).
\end{enumerate}

\section{Limitations}
\label{sec:limitations}
%
%We conclude by discussing the limitations of our work, clarifying the scope within which our conclusions hold and which extensions we leave to future work.

\mypar{Limited Performance of Dynamic Analysis.}
As previously observed~\cite{ponte2025slifer, dambra2023decoding}, dynamic analysis is less accurate than static analysis, and improving its effectiveness remains a challenging and still insufficiently explored research direction~\cite{kaya2025ml}.
In fact, Nebula~\cite{trizna2024nebula} requires large training sets and days of emulation; virtualization would yield richer traces but at considerably higher cost~\cite{kuchler2021does}.
Nevertheless, to the best of our knowledge, Nebula remains one of the best-performing architectures for dynamic malware detection~\cite{trizna2024nebula, ponte2025slifer}, which makes its inclusion in a Compound AI System representative of what a practitioner could realistically deploy today.
The limits we report for the dynamic level should therefore be read as limits of the current state of the art, not as artifacts of our choice of model.

\mypar{Only Static Adversarial EXEmples.}
Our robustness evaluation relies exclusively on attacks against static models, without targeting the dynamic level within the proposed threat models.
As discussed in \autoref{sec:robustness}, this reflects the current state of available tools: the relevant techniques either lack a public implementation~\cite{rosenberg2017bypassing, rosenberg2020query} or depend on outdated and unmaintained platforms~\cite{digregorio2024tarallo} difficult to set up and execute, in contrast to static attacks, for which established open-source solutions exist~\cite{demetrio2021secml}.
This choice is also consistent with our own findings: adversarial EXEmples that evade the static level are rarely stopped by the dynamic one (\ref{THM13}), so attacking the static level is where the attacker's effort is best spent.
Extending the threat models to manipulations of runtime behavior would enlarge the attack surface we measure, and we expect our robustness figures to be optimistic in that respect.

\mypar{Missing End-to-End Attacks.}
We did not compute query-based attacks~\cite{demetrio2021functionality, song2022mab} that optimize directly against the output of the Compound AI System; we instead attacked its first levels and transferred the resulting adversarial EXEmples to the complete pipeline.
As anticipated in \autoref{sec:robustness}, a query-based attack would require waiting for the emulator to terminate at every iteration of the optimization, making the process prohibitively slow for the number of samples and configurations considered here.
Transfer attacks provide a lower bound on attack effectiveness, and hence an upper bound on the robustness of the systems under test: an end-to-end attacker would be at least as effective as the one we model, which strengthens rather than weakens our comparative conclusions across systems.

\mypar{Robustness to Packing and Obfuscation.}
We did not explicitly evaluate detection performance under packing and obfuscation, nor did we control for their presence in the data.
The Speakeasy dataset~\cite{trizna2022quo} was collected in the wild by a security vendor and its authors report that packed samples are included, so such techniques are represented in our evaluation; we rely on this assessment without quantifying their prevalence.
%A dedicated study of how packing interacts with each level of a Compound AI System, and in particular with signature matching, is left to future work.

\mypar{Limited Data Sources.}
While large-scale~\cite{anderson2018ember, harang2020sorel} and more recent~\cite{joyce2025ember2024} datasets are available, they require paid access to VirusTotal\footnote{\url{https://www.virustotal.com}} to retrieve the corresponding programs, which we need to run the attacks and the emulation.
Our methodology is nonetheless independent of the specific data source and can be replicated on any dataset providing the raw executables.

\mypar{Temporal Drift.}
We did not explicitly address temporal drift, i.e., the shift in data distribution over time.
Unlike other domains~\cite{pendlebury2019tesseract}, establishing when a sample was first observed in the wild would require either vendor telemetry or VirusTotal access, neither viable here.
Our splits preserve a temporal ordering (Speakeasy training/test sets collected in January/April 2022, Chocolatey goodware in 2024) but this is not a substitute for an evaluation over a longer observation window.

\section{Future Work and Conclusions}
\label{sec:conclusions}

\mypar{Future Work.}
Our results identify the dynamic level as the weakest component of the pipeline, and improving it is our primary direction for future work: we will investigate richer representations of behavioral data, as well as alternative emulators and virtualization approaches that trade execution speed for fidelity~\cite{kuchler2021does}.
In parallel, we will extend our treatment of Compound AI Systems to the \emph{malware classification} task, i.e., attributing a sample to the family it belongs to, so that the same system can reliably both detect and characterize threats.
On the attacker side, we plan to broaden the threat models along two axes.
First, we will study techniques that evade dynamic detection directly, and the tampering of training data through poisoning attacks~\cite{biggio2018wild, cina2023wild}.
Second, as discussed in \autoref{sec:adv_evaluation}, an attacker who evades both the signature and the static level forces the system to fall back on emulation, inflating inference time.
We will investigate whether this behavior can be exploited systematically to mount \emph{Denial-of-Service} (DoS) attacks that overload the dynamic level, turning a robustness weakness into an availability one.

\mypar{Conclusions.}
In this work, we advocate for a Compound AI System paradigm in which multiple ML components cooperate and are complemented by classical detection techniques, moving beyond the evaluation of standalone models.
We pair this paradigm with system-level threat models that consider attackers with increasing knowledge of the deployed components, overcoming the limits of isolated robustness evaluations.
Together, these two contributions enable an explicit analysis of the trade-offs among detection performance, computational requirements, and robustness.
To this end, we implemented \obelisk, a Compound AI System whose levels reduce and refine the data forwarded from one stage to the next, and we compared several of its configurations, each adopting a different filtering policy at training and/or test time, against a standard version that applies no filtering.
Our experiments show that filtering policies substantially reduce training cost and improve responsiveness while sacrificing little detection performance. 
%(\ref{THM1},\ref{THM3}).
%
%whereas the unfiltered systems require more than $100$ hours of training (\ref{THM2}) and over $4$ seconds per sample at inference (\ref{THM4}).
%
Our four threat models further show that better-informed attackers craft adversarial EXEmples that transfer more effectively to the complete system, and that these attacks degrade responsiveness as a side effect.
The level-wise analysis explains where the system's robustness comes from, and where it is lost: signature-based components play a critical defensive role that collapses once the signatures are disclosed; overly aggressive filtering lets attackers bypass the pipeline by evading the static level alone; and dynamic analysis contributes to defense only marginally, and only when trained on the full dataset.
Efficiency gains and robustness losses thus stem from the same mechanism, which is precisely the trade-off a practitioner must resolve. We translate this into four deployment guidelines that map operational constraints to the best configuration.
Our work is a first step toward a more realistic evaluation of Compound AI Systems, narrowing the gap between industrial practice and academic research.

\section*{Acknowledgements}
This project has been partially funded by FISA-2023-00128 ``InfoAICert" funded by the MUR program “Fondo Italiano per le Scienze Applicate”, by SERICS (PE00000014) and by PNRR MUR Project (PE0000013) "Future Artificial Intelligence Research (FAIR)", funded by the European Union – NextGenerationEU, CUP J33C24000420007.

\bibliography{biblio}
\bibliographystyle{IEEEtran}

\end{document}